\documentclass[fleqn]{2017SCGE}
\usepackage[T1]{fontenc}
\usepackage{ae,aecompl}
\usepackage{enumerate,mathrsfs}
\usepackage{graphicx}	
\usepackage{amsmath}	
\usepackage{amssymb}	
\usepackage{ulem}
\usepackage{float}
\usepackage{epsfig}
\usepackage{color}
\usepackage{multirow}

\newcommand{\Rmnum}[1]{\expandafter\@slowromancap\romannumeral #1@}

\newcommand{\degree}{^{\circ}}
\newcommand{\sech}{\,\mathrm{sech}}
\newcommand{\csch}{\,\mathrm{csch}}

\newcommand\aap{A\&A}                
\newcommand\aplett{Astrophys.~Lett.} 
\newcommand\apj{ApJ}                 
\newcommand\apjl{ApJ}                
\newcommand\cjaa{Chinese J.~Astron. Astrophys.} 
\newcommand\mnras{MNRAS}             
\newcommand\nat{Nature}              
\newcommand\pasa{Publ. Astron. Soc. Australia}  

\begin{document}

\ensubject{subject}

\ArticleType{Article}
\SpecialTopic{SPECIAL TOPIC: FAST special issue}
\Year{2017}
\Month{January}
\Vol{60}
\No{1}
\DOI{10.1007/s11432-016-0037-0}
\ArtNo{000000}
\ReceiveDate{January 11, 2016}
\AcceptDate{April 6, 2016}

\title{The Radiation Structure of PSR B2016$+$28 Observed with FAST}{The Radiation Structure of PSR B2016$+$28 Observed with FAST}

\author[1,2]{Jiguang LU}{{lujig@nao.cas.cn}}%
\author[1,2]{Bo PENG}{pb@nao.cas.cn}
\author[3]{Renxin XU}{}
\author[1]{Meng YU}{}%
\author[4,1]{\\Shi DAI}{}
\author[1]{Weiwei ZHU}{}
\author[1,5]{Ye-Zhao YU}{}
\author[1]{Peng JIANG}{}
\author[1]{Youling YUE}{}
\author[1,5]{Lin WANG}{}
\author[]{\\ FAST Collaboration}{}

\AuthorMark{Lu J G}

\AuthorCitation{Lu J G, Peng B, Xu R X, et al}

\address[1]{CAS Key Laboratory of FAST, National Astronomical Observatories, Chinese Academy of Sciences, Beijing 100101, China}
\address[2]{Guizhou Radio Astronomy Observatory, Chinese Academy of Sciences, Guiyang 550025, China}
\address[3]{School of Physics and Kavli Institute for Astronomy and Astrophysics, Peking University, Beijing 100871, China}
\address[4]{CSIRO Astronomy and Space Science, Australia Telescope National Facility, Box 76 Epping NSW 1710, Australia}
\address[5]{College of Astronomy and Space Sciences, University of Chinese Academy of Sciences, Beijing 100049, China}


\abstract{
With the largest dish Five-hundred-meter Aperture Spherical radio Telescope (FAST),
both the mean and single pulses of PSR B2016$+$28,
especially including the single-pulse structure, are investigated in detail in this study.
The mean pulse profiles at different frequencies can be well fitted in a conal model,
and the peak separation of intensity-dependent pulse profiles increases with intensity.
The integrated pulses are obviously frequency dependent
(pulse width decreases by $\sim\,20\%$ as frequency increases from 300\,MHz to 750\,MHz),
but the structure of single pulses changes slightly
(the corresponding correlation scale decreases by only $\sim\,1\%$).
This disparity between mean and single pulses provides independent
evidence for the existence of the RS-type vacuum inner gap,
indicating a strong bond between particles on the pulsar surface.
Diffused drifting sub-pulses are analyzed. The results show that
the modulation period along pulse series ($P_3$) is positively correlated to the separation
between two adjacent sub-pulses ($P_2$).
This correlation may hint a rough surface on the pulsar,
eventually resulting in the irregular drift of sparks.
All the observational results may have significant implications
in the dynamics of pulsar magnetosphere and are discussed extensively in this paper.
}



\maketitle


\begin{multicols}{2}
\section{Introduction}
\label{sect:intro}

Pulsar magnetospheric activity depends on the binding energy of the particles on the stellar surface,
and it is thus relevant to the inner structure~\cite{lu18}.
It is generally believed that an RS-type vacuum inner gap~(\cite{rude75}; hereafter, RS) is responsible for sub-pulse drifting,
which could be quantitatively understood in a partially
\Authorfootnote

\noindent screen scenario for a conventional neutron star~\cite{gil03}
but could alternatively be explained using a bare strange star model~\cite{xu99}.
It is then evident that single-pulse observation may be the key to probe the pulsar surface,
and a microscopic structure analysis of single pulse is necessary.
Observation of strong pulsars with a large telescope can help in the study of the fine structure of individual pulses,
and the Five-hundred-meter Aperture Spherical radio Telescope (FAST) may be useful for this purpose ~\cite{peng00a,peng00b}.

PSR B2016$+$28, with flux density $\sim$314\,mJy at 400\,MHz~\cite{lori95},
is one of the brightest radio pulsars in the northern sky.
Further, the radiation properties of this pulsar are not yet clear; hence, this pulsar is selected for the study.

It has period $P_0\sim$0.558\,s~\cite{hobb04}
and dispersion measure DM$\sim$14.2\,$\mathrm{cm^{-3}\,pc}$~\cite{stov15}.
Observation at 430\,MHz and 1720\,MHz exhibited a not-well-resolved
double-humped pulse profile~\cite{oste77}.
Later, Rankin (1983) \cite{rank83} observed this pulsar at 430, 2700, and
4900\,MHz, and obtained similar results.
It was found that the profile of PSR B2016$+$28 widens
at progressively lower frequency and ultimately bifurcates.
Rankin (1983) \cite{rank83} classified such profiles as the conal-single type,
which indicates that the two humps of the pulse profile arise from one radiation cone.
Whereas, Weltevrede, Edwards \& Stappers (2006) \cite{welt06} analyzed the longitude resolved fluctuation spectrum (LRFS)
of this pulsar at wavelength 21\,cm ($\sim$1400\,MHz), and the much stronger
``slow'' drift mode in the leading part than in the trailing part of the pulses implied that
the two components in its pulse profile are distinct.

Similar to other pulsars, PSR B2016$+$28 exhibits the drifting
sub-pulse phenomenon.
Notably, its drift pattern is irregular.
Drake \& Craft (1968) \cite{drak68} first found that the sub-pulses of PSR B2016$+$28 have multiple drift rates.
Taylor, Manchester \& Huguenin (1975) \cite{tayl75} showed that the modulation period along the pulse series, $P_3$,
varies between $3P_0$ and $12P_0$.
Then, Rankin (1986) \cite{rank86} adjusted its upper limit to $15P_0$.
Furthermore, Weltevrede, Edwards \& Stappers (2006) \cite{welt06} and Weltevrede, Stappers \& Edwards (2007) \cite{welt07} demonstrated the continuous LRFS of
PSR B2016$+$28, and they identified the broad drifting features of this pulsar.
They also showed that $P_3$ of the ``slow'' drift mode is $\sim20P_0$.

The drifting feature of PSR B2016$+$28 also varies with frequency.
Oster, Hilton \& Sieber (1977) \cite{oste77} suggested that the radiation at 430 and 1720\,MHz have different
sub-pulse drift rates.
However, Naidu et al. (2017) \cite{naid17} obtained a contrasting result that the drift pattern is correlated across frequencies.
Additionally, Weltevrede, Stappers \& Edwards (2007) \cite{welt07} showed that the ``slow'' drift mode at 21\,cm band
vanishes at 92\,cm band ($\sim$330\,MHz).

In this study, to explain the aforementioned discrepancies,
we analyzed the radio emission (both mean and single pulses) of PSR B2016$+$28 observed with FAST.
In Section \ref{sect:observation}, the observation and data reduction approach are described.
In Section \ref{sect:profile}, we present an analysis of the mean pulses of PSR B2016$+$28 and
describe the beam structure of its radio radiation.
In Section \ref{sect:drift} and Section \ref{sect:evolution}, analyses of the sub-pulse drifting and
the evolution of individual pulses versus frequency, respectively, are presented.
We present the discussions in Section \ref{sect:discussion} and
summarize the results in Section \ref{sect:summary}.

\section{Observation and Data Reduction}
\label{sect:observation}

FAST is located in Guizhou, China, at longitude
$106.9\degree$\,E and
latitude $25.7\degree$\,N.
It was completed on 25th September 2016.
The aperture of the telescope is 500\,m and the effective aperture is about 300\,m.
FAST is in the test phase at the moment and it is able to track source over a period of time.
An ultra-wideband receiver with a bandpass 270-1620\,MHz was
used to take the observation.
%
%
The received signal was digitized with a sampling rate 1.8\,GHz, and was then filtered into
a low-frequency band 270-800\,MHz and a high-frequency band 1250-1620\,MHz~\cite{jian19}.
After processing by the filter bank in a Reconfigurable Open Architecture
Computing Hardware (ROACH), the data were recorded as the search mode 
PSRFITS format~\cite{hota04}.

In this work, the 270-800\,MHz data of PSR B2016$+$28 observed
on 12th September 2017 were used.
The total observation time is 22 minutes.
The data have a time resolution 0.2\,ms and a frequency resolution 0.25\,MHz.
In our work, the data were incoherently de-dispersed with
DM=14.1977\,$\mathrm{cm^{-3}\,pc}$~\cite{stov15}.
Then the de-dispersed data were folded with the Chebyshev predictor calculated by
TEMPO2~\cite{hobb06,edwa06} with the ephemeris provided by PSRCAT\footnote{see
http://www.atnf.csiro.au/research/pulsar/psrcat/} (version 1.56),
and the phase bin number was set to be 2048 bin/cycle.
The channels with distinct radio frequency interference were removed manually.
To study the pulse evolution with the frequency, the data in the band 275-775\,MHz were also
split into ten narrow bands with a bandwidth 50\,MHz and frequency-scrunched into single
pulse trains respectively.
The mean and single pulses of these ten narrow bands are analyzed below.

\section{Mean Pulse Profiles}
\label{sect:profile}

\begin{figure*}
  \centering
   \includegraphics[width=0.9\linewidth]{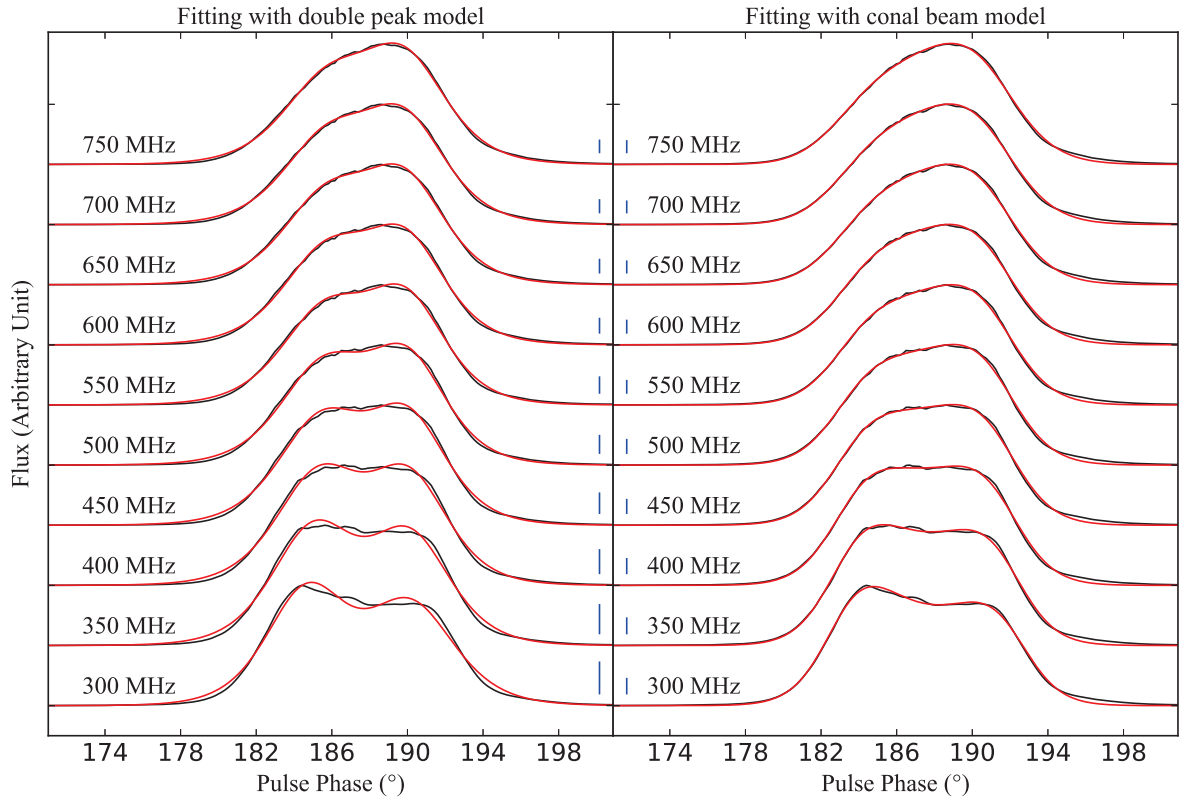}
   \caption{{\small The narrow band profiles are fitted with two square hyperbolic secant
   functions (left panel) and a conal beam function (right panel).
   The black curves are the observed pulse profiles (had been normalized) and the red curves
   are the fitting curves.
   The blue vertical line is 10 times of the rms (root-mean-square) of fitting residuals.
   Obviously, the conal beam function gets better fitting results.}}
   \label{profile_fit}
\end{figure*}

Integrated pulse profile of PSR B2016$+$28 can be treated as the conal-single type~\cite{rank83} or
two distinct components~\cite{welt06}, and here we tried to fit it with double peaks
or a conal beam.
In the double peaks fitting, the specific fitting function has following expression,
\begin{equation}
F_{\mathrm{double~peaks}}(\phi)=\sum^{2}_{i=1}F_{\mathrm{sech~square}}(\phi,A_i,\sigma_i,\phi_{0,i}),
\label{twopeaksfunction}
\end{equation}
where $\phi$ is the pulse phase of the components; $A_i$, $\phi_{0,i}$ and $\sigma_i$ are used to describe the shape of the $i$-th peak; $F_{\mathrm{sech~square}}$ is the square hyperbolic secant function,
\begin{equation}
F_{\mathrm{sech~square}}(\phi,A,\sigma,\phi_{0})=A\sech^2\left(\sqrt{\frac{2}{\pi}}\frac{\phi-\phi_{0}}{\sigma}\right),
\label{sechfunction}
\end{equation}
which is examined in Lu et al. (2016) \cite{lu16} to be the most appropriate function to describe
the shape of the pulse profile wings.
On the other hand, to fit profiles with a conal beam, the cone function discussed in Lu et al. (2016) \cite{lu16}
is adopted,
\begin{equation}
F_{\mathrm{conal~beam}}=F_{\mathrm{sech~square}}(\theta_{\mu},A,\sigma,\theta_{\mu0})\times[1+k\cdot(\phi-\phi_0)],
\label{conalfunction}
\end{equation}
where $\theta_{\mu0}$ is the angular radius of the conal beam, and $\theta_{\mu}$ is the angular distance between the magnetic axis and the radiation direction,
and it can be calculated with the following equation,
\begin{equation}
\cos\theta_{\mu}=\cos\alpha\cos(\alpha+\beta)+\sin\alpha\sin(\alpha+\beta)\cos(\phi-\phi_0),
\label{geometry}
\end{equation}
where $\alpha$ is the magnetic inclination angle and $\beta$ is the impact angle,
and in this work the values $\alpha=40.4\degree$\ and $\beta=6.2\degree$\ obtained
in Lyne \& Manchester (1988) \cite{lyne88} are adopted.
In Eq.~\ref{conalfunction},
$A$ is the amplitude of the function,
$\sigma$ is the angular thickness of the cone wall,
$k$ is the first order of the radiation intensity variation versus pulsar rotating phase,
which could be regarded as the asymmetry of conal component,
and $\phi_0$ is the pulse phase at the time when the line of sight, the
conal beam axis, and the rotation axis are coplanar~\cite{lu16}.

\begin{figure}[H]
  \centering
   \includegraphics[width=0.8\linewidth]{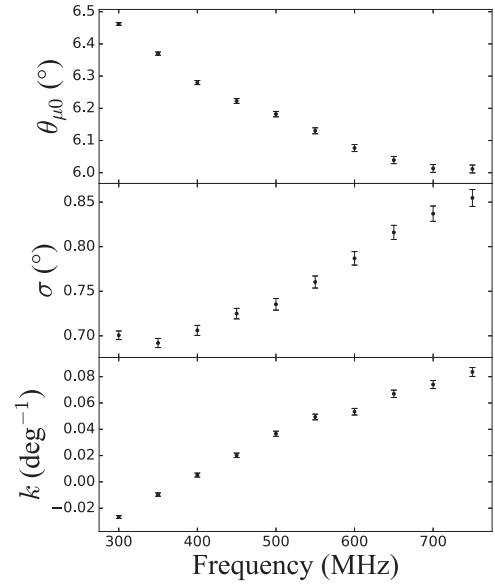}
   \caption{{\small Best-fitting parameters of the conal beam model vs. frequency.
   From top to bottom, the parameters are the conal radius ($\theta_{\mu0}$),
   the cone wall thickness ($\sigma$), and the conal asymmetry ($k$). }}
   \label{profile_fit_result}
\end{figure}

The fitting results are shown in Figure~\ref{profile_fit} (the Levenberg-Marquardt algorithm
is used to fit the profiles, and it is also adopted below).
Obviously, the conal beam function could fit the profiles better, especially at the low frequency.
As the distance between two peaks increases at the lower frequency,
the bridge region between two peaks could play an important role.
As shown in Lu et al. (2016)~\cite{lu16}, the bridge component will lead to bad
fitting results for two peaks model.
Whereas the two peaks in high frequency profiles are unresolved,
and those profiles could be well fitted by two peaks model.
Above fitting results indicates that the radiation of PSR B2016$+$28 is more likely generated
from a conal beam than two independent radiation components.
Thus, in the following work, the pulse phase would not be split into two
intervals while studying the single pulse.
The evolution of best fitting parameters with frequency is shown in Figure~\ref{profile_fit_result}.

The profile evolution properties which shown in Figure~\ref{profile_fit_result} can be compared with that of
PSR B1133$+$16 (Figure 9 in Lu et al. (2016) \cite{lu16}).
We found that radii of the conal beam ($\theta_{\mu0}$) of both PSR B2016$+$28 and PSR B1133$+$16
decrease with frequency.
But the thickness and asymmetry of the cone ($\sigma$ and $k$) shows different evolution direction.
Lu et al. (2016) \cite{lu16} shows that the thickness of the conal beam wall represents the particle energy dispersion.
Then the thickness evolution in Figure~\ref{profile_fit_result} naturally implies the particle energy dispersing process.
Obviously, the dispersing process in PSR B2016$+$28 dose not show a cut-off like the Figure 13 in Lu et al. (2016) \cite{lu16},
and this may be resulted in the frequencies of adopted PSR B2016$+$28 data are too low to exhibit the cut-off.
However, the different evolution direction of asymmetry may be confusing.
The inclination angle and impact angle of PSR B1133$+$16 are $51.3\degree$
and $3.7\degree$, respectively.
It seems that both pulsars have similar inclination angles (both are less than 90$\degree$) and impact angles (both are positive), and they should also have similar radiation beams for similar period of them (the period of PSR B1133+16 is $\sim$1.188\,s).
But the asymmetry evolution tendencies of them are the opposite.
One probability is that the values $\vec{\Omega}\cdot\vec{B}$ of
two pulsars have different sign, and this may lead to difference between their beams.

\begin{figure}[H]
  \centering
   \includegraphics[width=\linewidth]{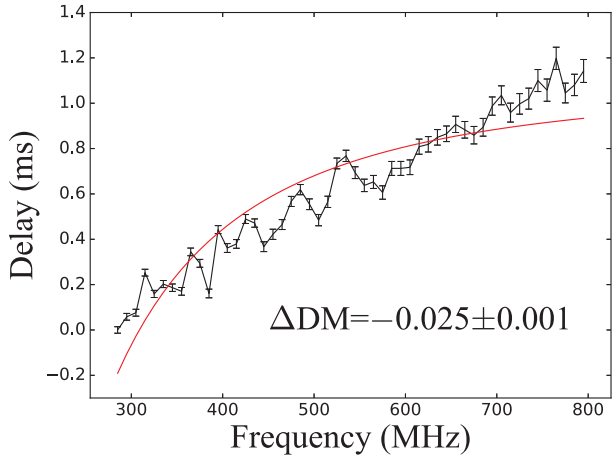}
   \caption{{\small Evolution of the conal beam center position.
   The ordinate is the time delay of conal beam center relative
   to that at 285\,MHz.
   The evolution is fitted by a correction of DM, and it is shown by the black curve.}}
   \label{dm_fit}
\end{figure}

The evolution of parameter $\phi_0$ is not shown in Figure~\ref{profile_fit_result},
because $\phi_0$ just has a relative value.
But the evolution of $\phi_0$ would influence the measure of DM.
In the previous work, the value of DM is always obtained by aligning the profile at different
frequencies with a profile template, and this process is really based on the
assumption that the pulse profiles are unified across frequencies.
However, the profiles themselves do not keep with frequency.
Thus, The evolution of the beam may bring small derivation between calculated value and actual value of DM.
On the other hand, $\phi_0$ is the phase of conal beam center,
and it may supply a more accurate method to calculate DM.
Accordingly, the 280-800\,MHz data are split into 32 narrow bands and the profiles in these bands
are fitted with Eq.~\ref{conalfunction} to obtain the relative $\phi_0$.
Then these relative phase movements are fitted with a DM correction, which is shown in Figure~\ref{dm_fit}.
From the figure, the real DM should be a little smaller than the value we used.

\section{Single Pulse: Drifting Sub-pulse}
\label{sect:drift}

\begin{figure}[H]
  \centering
   \includegraphics[width=\linewidth]{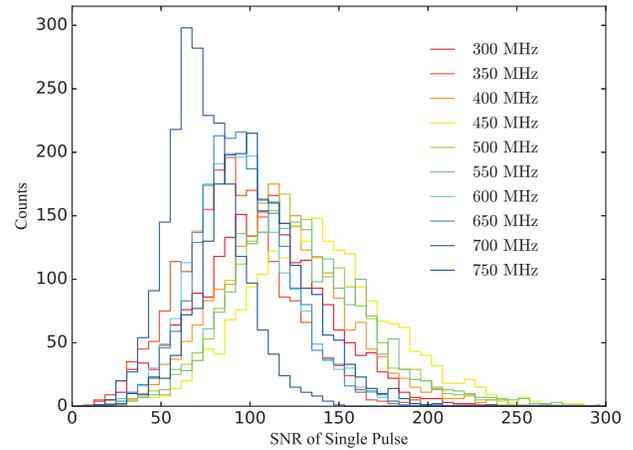}
   \caption{{\small The statistics of single pulse SNR, and narrow bands with different frequencies are marked with different colors.}}
   \label{single-pulse-snr}
\end{figure}

Before analyzing the single pulse, the statistics of single pulse signal-to-noise ratio
(SNR, calculated by dividing the mean intensity on every bin in pulse-on region, i.e., bins
within phase range $170\degree-205\degree$, by the rms intensity on pulse-off region)
at each narrow band is shown in Figure~\ref{single-pulse-snr}.
In the figure, it could be found that the SNR is large enough to do deep analysis.

\subsection{Correlation Analysis}
\label{sect:corr_analysis}

\begin{figure}[H]
  \centering
   \includegraphics[width=\linewidth]{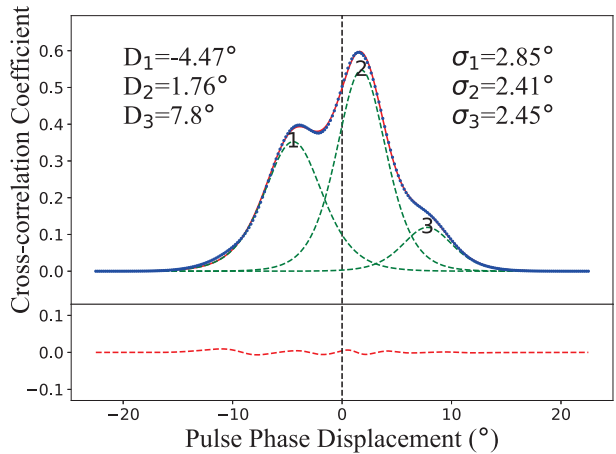}
   \caption{{\small The mean cross-correlation coefficients of successive
   pulses are shown as blue dots, and they seems being composed of 3 components.
   It is tried to fit the coefficients with 3 square hyperbolic secant peaks,
   and the fitting result is plotted as red curve.
   The 3 green dashed curves are the 3 peaks, respectively,
   and the red dashed curve shows the fitting residuals.
   The centric position $D_i$ and width $\sigma_i$ of 3 peaks are also marked.}}
   \label{cross-correlation}
\end{figure}

To determine the approximate drift rate, the cross-correlation of successive
pulses is analyzed, as Sieber \& Oster (1975) \cite{sieb75} did to generate their Figure 4.
The cross-correlation result is shown in Figure~\ref{cross-correlation} as blue dots,
and the curve composed by them seems having 3 components.
Then we try to fit this curve with 3 square hyperbolic
secant peaks, and determine the centric position $D_i$ and
width $\sigma_i$ ($i=1-3$) of these 3 components.

\begin{figure*}
  \centering
   \includegraphics[width=0.4\linewidth]{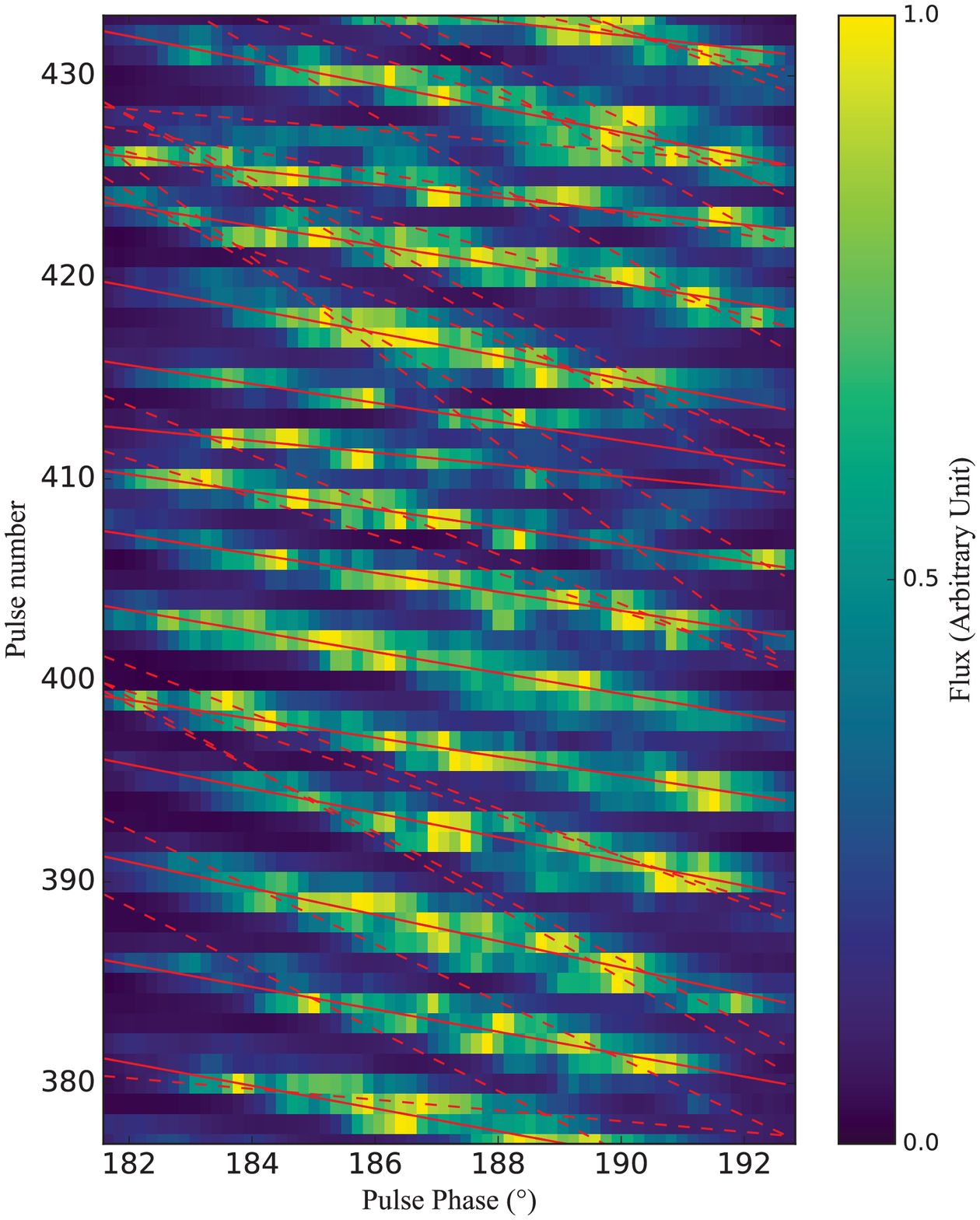}
   \includegraphics[width=0.4\linewidth]{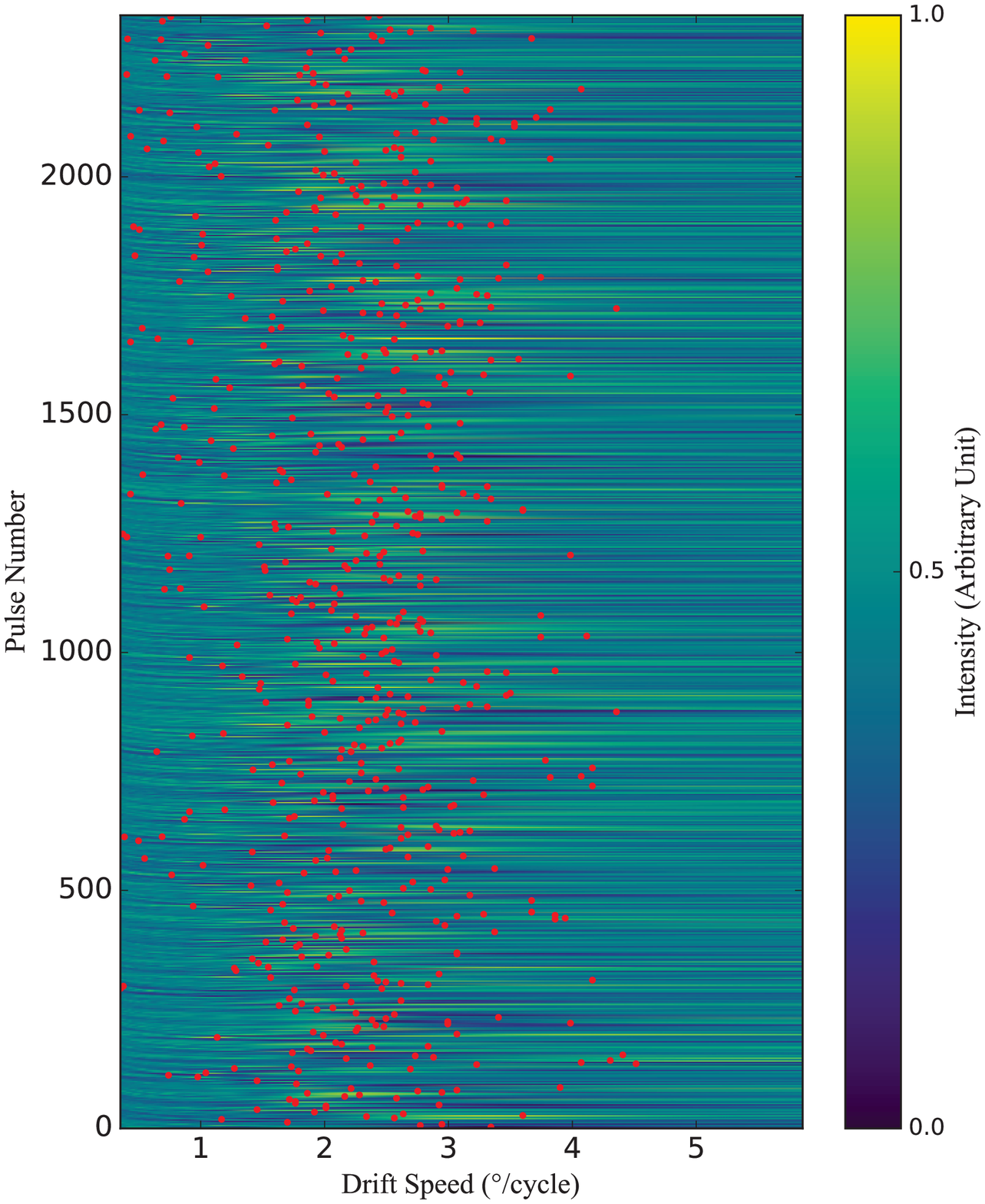}
   \caption{{\small Part of single pulses are shown in left panel,
   and the drifting sub-pulses are marked as red solid lines.
   The red dashed lines are fake drift tracks (both solid lines and dash lines corresponds to the red points in right panel).
   The flux variation at different drift rates is shown in right panel,
   and each maximum point (red point) corresponds to a sub-pulse track.}}
   \label{sub-pulse_drift}
\end{figure*}

\begin{figure}[H]
  \centering
   \includegraphics[width=0.9\linewidth]{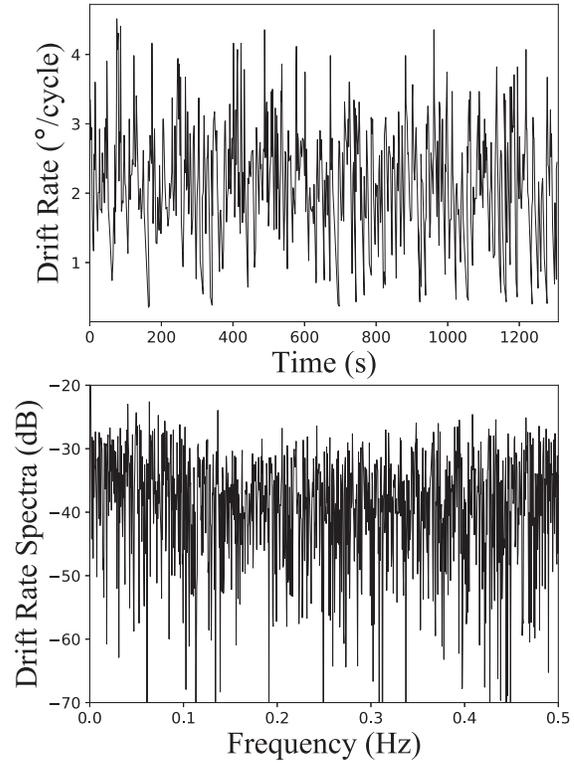}
   \caption{{\small The drift rates of the extracted 622 sub-pulse tracks variation versus time
   (top panel) and the variation spectra of the drifting rate (bottom panel).}}
   \label{drift_rate}
\end{figure}

Cross-correlation method could show the correlation between adjacent single pulses,
and the maximum point of the correlation result should be the relative shift between them.
Thus the averaged cross-correlation between every two adjacent single
pulses will show the mean correlation result, as shown in Figure~\ref{cross-correlation}.
Then the separation between the peaks 1 and 2 or 2 and 3 in averaged cross-correlation result can roughly represent the mean separation between adjacent sub-pulses, i.e., $P_2$.
The peak position in the averaged result is the mean relative shift between two adjacent single pulses, i.e., mean drift phase in one period (mean drift rate).
The drift rate estimated here could be used to estimate the range of
real drift rate which is used in the analysis below (such as in determining
the range of abscissa in right panel of Figure \ref{sub-pulse_drift}).
However, there are three peaks in Figure \ref{cross-correlation}, and only one could represent the real drift
(the other two peaks are fake drift bands with a same $P_2$ and different $P_3$).
The real drift rate need to be determined.
As Sieber \& Oster (1975) \cite{sieb75} showed, the different components of cross-correlation result
represent different drifting bands, and it is hard to judge the real drift direction of pulsar.
In fact, that ambiguity for coherent drifters (pulsar with stable and fixed $P_3$,
like PSR B0943+10) may not be a problem for diffuse drifters (pulsar with broad
drift feature).
For coherent drifters, the sub-pulse would drift regularly,
and drift rate is stable for every sub-pulse.
Then, in cross-correlation result (similar to the green dash curves in Figure~\ref{cross-correlation})
the peaks corresponding to real and fake drift bands would have the same width.
Thus the real and fake drift bands are hard to differentiate.
However, for the sub-pulses of a diffused drifter like PSR B2016+28, the drift rate for a specified sub-pulse is stable, and for different sub-pulses, the drift rates would be different.
Therefore, in the cross-correlation result, the peak representing the fake band should be wider than that representing the real band.
Then the narrowest peak could be treated as the real drift band.
As shown in Figure~\ref{cross-correlation}, the narrowest component is Component 2,
thus it is chosen as the real drift direction in this paper.
%
%
In fact, Component 2 is also chosen to be analyzed in previous
works~\cite{oste77,rank86,welt06,welt07,naid17}.
Fortunately, Component 2 is also the strongest component, and it means that
Component 2 is more convenient to be analyzed.

\begin{figure}[H]
  \centering
   \includegraphics[width=\linewidth]{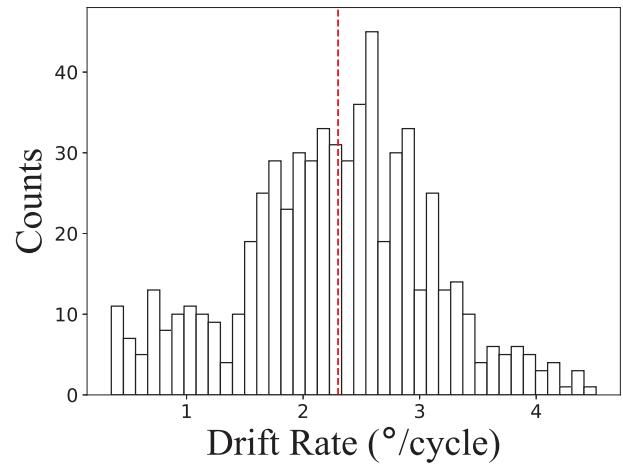}
   \caption{{\small The statistics of the extracted 622 drifting rates.
   The red dashed line marks the division point of the slower and faster
   drift sub-pulses for Figure~\ref{drift_rate_profiles}.}}
   \label{drift_rate_statistics}
\end{figure}

\subsection{Drift Rate of Sub-pulse}
\label{sect:drift_rate}

\begin{figure}[H]
  \centering
   \includegraphics[width=\linewidth]{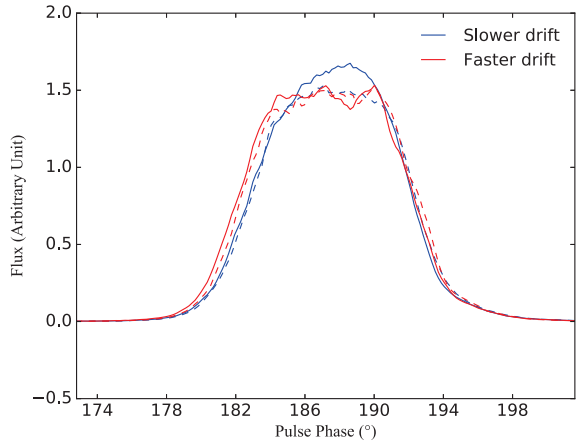}
   \caption{{\small The profiles of slower (red curves) and faster (blue curves)
   drift sub-pulses, and the division point is shown in Figure~\ref{drift_rate_statistics}.
   The solid (dashed) curves represent the profiles of first (second) half of sub-pulses.}}
   \label{drift_rate_profiles}
\end{figure}

There are altogether 2347 continuing single pulses in the pulse trains.
Part of single pulses are shown in Figure~\ref{sub-pulse_drift} (left panel), and the drifting sub-pulse
can be identified clearly.

To determine the drift rate of each sub-pulse, the flux variation at different
drift rates is considered.
For a given drift rate, the flux of pulse-on region (between pulse phase 170$\degree$ and 205$\degree$)
is added along the specified drift rate to form a flux variation series
(i.e. phase-scrunching left panel in Figure~\ref{sub-pulse_drift} along specified slope).
In the left panel of Figure 9, the sub-pulses could be find drifting along pulse tracks, and the drift rates of each sub-pulse is right the slope of that train track.
Thus, for each drift rate, there is a corresponding inclination direction
on the left panel of Figure \ref{sub-pulse_drift}, and the left panel of Figure \ref{sub-pulse_drift}
could be compressed along this direction (sum the 2-D time-domain pulse flux
array along this direction to form a 1-D flux array).
With this method, the 2-D time-domain pulse flux array could correspond
to a flux variation series for different drift rates,
and it is shown in the right panel of Figure \ref{sub-pulse_drift} (every point on it could be corresponding
to an inclining line on the left panel).
The drift rate in the range 0.3-6.0\,$\degree$/cycle is chosen to be analyzed
based on the result obtained in Section~\ref{sect:corr_analysis}
(drift band 2 in Figure~\ref{cross-correlation} is real).
The flux variation at different drift rates is also shown in Figure~\ref{sub-pulse_drift} (right panel).
In the figure, each maximum point (marked as red points in the right panel of Figure~\ref{sub-pulse_drift})
corresponds to a possible drifting sub-pulse track.
With the drift rate and pulse number of these maximum points, the corresponding
drift sub-pulse track can be determined.
These tracks are also marked on the left panel of Figure~\ref{sub-pulse_drift} as
red lines.
On the left panel, for some lines there is no obvious sub-pulse track along it.
It may be caused by the fake band like the band A in Fig. 1 in Sieber \& Oster (1975)\cite{sieb75},
and also it may be caused by a weak sub-pulse-train
(because there are few weak pulses as shown in Figure 4).
Then the fake tracks are distinguished from the real ones empirically.
and these tracks are marked as dashed lines.
Compareing with the conventional method used in Leeuwen et al. (2002) \cite{leeu02},
this method have two advantages.
Firstly, the analysis is independent on the shape of sub-pulses
(the shape of the sub-pulse need not to have a Gaussian-like shape).
Secondly, the sub-pulse trains are allowed to intersect each other.

Totally, 622 sub-pulse tracks are manually extracted as real ones
from all tracks, and they are analyzed below.
The drift rates of these 622 sub-pulse tracks variation versus time and
the variation spectra of them are shown respectively in the
top panel and bottom panel of Figure~\ref{drift_rate}.
From the figure, the drift rates of sub-pulses have no distinct mode
like PSR B1819-22~\cite{sery11}, and also no period and quasi-period
like PSR B0943+10~\cite{desh01}.
On the other hand, the statistics result of these drifting rates is shown in
Figure~\ref{drift_rate_statistics}, and it also shows a diffused distribution.
We also tried to add slower drift sub-pulses (drift rate smaller than
$2.3\degree$/cycle; this dividing point is just an experimental value and is marked
as a vertical red dashed line in
Figure~\ref{drift_rate_statistics}) and faster drift sub-pulses (drift rate higher
than $2.3\degree$/cycle) to form profiles respectively.
For this analysis, the whole pulse trains are split into two halves with equal
pulse numbers for comparison.
The results are shown in Figure~\ref{drift_rate_profiles}.
In the figure, the blue (red) curves represent the profiles of slower (faster) drift sub-pulses,
while the solid (dashed) curves represent the profiles of first (second) half of sub-pulses.
From the figure, the profile of slower drift sub-pulses is stronger at leading wing, and there is
no distinct difference in other phase region.

\begin{figure}[H]
  \centering
   \includegraphics[width=\linewidth]{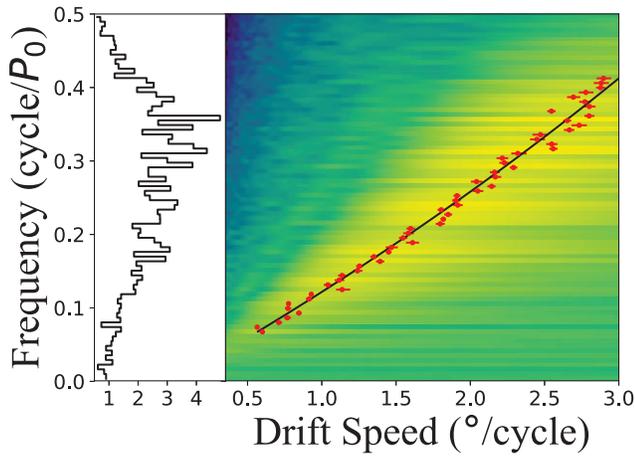}
   \caption{{\small The flux modulation at different drift rates (right panel) and
   the total modulation spectra (left panel).
   The red points mark the maximum points along drift rate at different
   modulation frequency, and the black curve is the fitting result.}}
   \label{drift_rate_f3}
\end{figure}

In Figure \ref{drift_rate}, the bottom panel shows the modulation spectra
of the drift rate sequence, and no period could be obtained.
However, in the right panel of Figure \ref{sub-pulse_drift}, at a specified drift rate,
the flux series seems to have an indistinct modulation,
and the modulation period seems to be larger at higher drift rate region.
It is motivated for carrying out the Fourier transform along vertical
axis for right panel of Figure \ref{sub-pulse_drift},
and the result is shown in the right panel in Figure \ref{drift_rate_f3}.
The modulation period along pulse trains is just the definition of $P_3$,
thus the ordinate of Figure \ref{drift_rate_f3} is right the frequency corresponding to $P_3$.
Thus Figure \ref{drift_rate_f3} could show the relation between $P_3$ and drift rate.
The spectra of modulation frequency is shown in the left panel of the figure,
and it has broad distribution just like the result obtained in Weltevrede, Stappers \& Edwards (2007) \cite{welt07}.
From the figure, the modulation frequency varies with drift rate.
The maxima of the Fourier spectra at each modulation frequency has been marked as
red point and error bar in Figure~\ref{drift_rate_f3}, and it seems that there is some
relation between drift rate $D$ and modulation frequency $F_3$.
Assume that this relation have the following form,
\begin{equation}
D=\frac{F_3}{k_d F_3+b_d},
\label{f2f3fit1}
\end{equation}
where $k_d$ and $b_d$ are the parameters to be fitted.
The parameters which give best fitting results are
$k_d=20\pm3$, $b_d=41\pm1$\,cycle/$P_0$, and
the best fitting curve is also plotted in the figure.
In fact, for a coherent drifter, the drift rate is right the quotient of
$P_2$ and $P_3$.
Then the relation between $P_2$ and $P_3$ could be obtained via fitting the
points in Figure~\ref{drift_rate_f3},
and this relation could be expressed as following,
\begin{equation}
P_2=\frac{P_3}{\gamma_d P_3+\beta_d},
\label{p2p3fit1}
\end{equation}
where $\gamma_d=74\pm1$ and $\beta_d=20\pm3$\,s,

\begin{figure}[H]
  \centering
   \includegraphics[width=\linewidth]{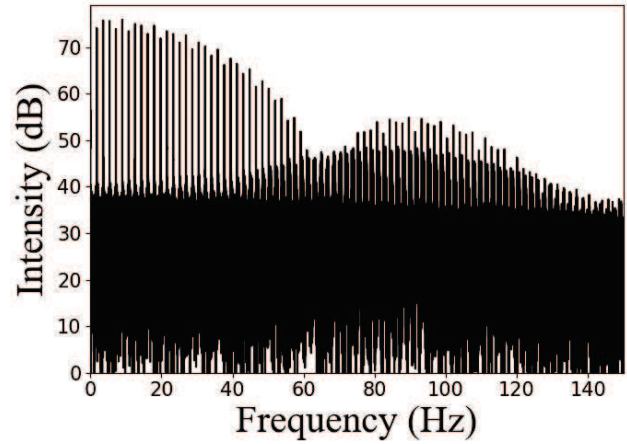}
   \caption{{\small The unfolded fluctuation spectra of time-domain data of pulses.}}
   \label{fft_result}
\end{figure}

\begin{figure}[H]
  \centering
   \includegraphics[width=\linewidth]{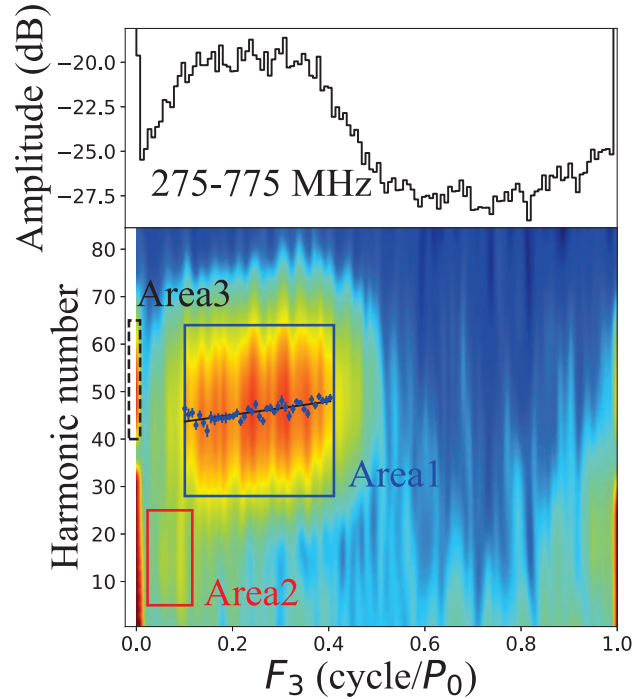}
   \caption{{\small The folded fluctuation spectra of time-domain data
   of pulses (bottom panel) and the intensity distribution with different $P_3$.
   In the bottom panel, three distinct areas are marked.
   In Area 1, the maximum points along vertical axis are marked
   as blue points and error bars, and the fitting curve is also plotted.)}}
   \label{harmonic_analysis}
\end{figure}

\subsection{Harmonic Analysis}

\begin{figure}[H]
  \centering
   \includegraphics[width=\linewidth]{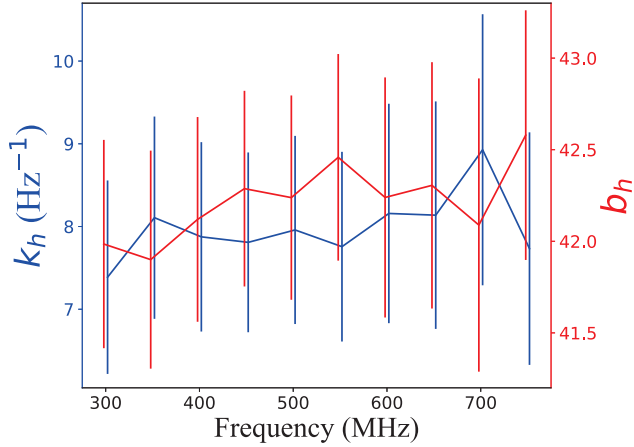}
   \caption{{\small The best-fitting parameters $k_h$ and $b_h$ of Eq.~\ref{f2f3fit2}
   versus frequency.
   It seems that $b_h$ shows a little increasing with frequency.}}
   \label{area1_fit_result}
\end{figure}

Fourier transformation is done for the frequency-scrunched data sequence to
obtain the fluctuation spectra (as shown in Figure~\ref{fft_result}).
Naturally, there are a series of peaks at fundamental and multiple frequencies of pulses.
Then the fluctuation spectra is folded with the pulsation frequency just like the
method to obtain Figure 4 in Deshpande \& Rankin (2001) \cite{desh01}.
The folded spectra are shown in Figure~\ref{harmonic_analysis}, where the multiple
frequencies of pulses have been move to the center of the figure.

In Figure~\ref{harmonic_analysis}, the abscissa is labeled $F_3$, because
it happens to correspond to $1/F_3$.
For a simple example of a pulse series with period $p_0$ and a Gaussian shape profile
$\frac{1}{\sqrt{2\pi}\sigma}\exp\left(-\frac{t^2}{2\sigma^2}\right)$,
if it also has sine shape sub-pulse and $P_2=p_2$,
its fluctuations could be expressed as following,
\begin{equation}
f=\left[\frac{1}{2\pi}+\frac{1}{\pi}\sum^\infty_{n=1} \exp\left(-\frac{2\pi^2n^2\sigma^2}{p_0^2}\right)
\cos\left(\frac{2\pi n}{p_0}t\right)\right]\cos^2\left(\frac{\pi}{p_2}t\right),
\label{simulatedf}
\end{equation}
where the expression in the square bracket is the approximative Fourier series
of Gaussian in the period of $p_0$, the square of cosine function is
the shape of sinusoidal sub-pulse with a period of $p_2$.
In the fluctuation spectra, it should have peaks at $\frac{n}{p_0}$
(the fundamental and multiple frequencies of pulses, $n$ is the frequency multiplication of pulse)
and $\frac{n}{p_0}+\frac{1}{p_2}$ (the frequency due to drifting sub-pulses).
Obviously, the drift rate of sub-pulse $d$ ($d=f_3/f_2$) satisfies $p_0\equiv dp_0~(\mathrm{mod}\,p_2)$,
then it could be obtained $f_2\equiv df_2~(\mathrm{mod}\,f_0)$,
where $f_2$, $f_3$ and $f_0$ are the frequencies corresponding to $p_2$, $p_3$ and $p_0$.
In other words, the frequency due to drifting sub-pulses
$\frac{n}{p_0}+\frac{1}{p_2}=nf_0+f_2$
could be expressed as $hf_0+f_3$ ($h$ is the harmonic number), i.e.,
the horizontal axis of Figure~\ref{harmonic_analysis} should be $F_3$.
In the top panel of Figure~\ref{harmonic_analysis}, the broad distribution of $F_3$
could also be found.

In the bottom panel of Figure~\ref{harmonic_analysis}, three distinct areas are marked.
Among them, Area 1 is the drift component analyzed in Taylor, Manchester \& Huguenin (1975) \cite{tayl75} and Rankin (1986) \cite{rank86};
Area 2 is the ``slow'' drift mode component mentioned in Weltevrede, Edwards \& Stappers (2006) \cite{welt06};
Area 3 is the multiple frequencies of pulses, also the ripples of the
profile spectra.

In Area 1, the harmonic number of maximum points along ordinate $H$ (which have been
marked as blue points and error bars) seem varying versus $F_3$.
It is tried to fit this variation with a linear relation,
\begin{equation}
H=k_h F_3+b_h,
\label{f2f3fit2}
\end{equation}
where $k_h$ and $b_h$ are the parameters to be fitted.
The best-fitting parameters are $k_h=14\pm2\,P_0$/cycle,
$b_h=42.3\pm0.5$.
We also fit the variation of $H$ in each narrow band with 50\,MHz bandwidth,
and the best-fitting parameters $k_h$ and $b_h$ are shown in Figure~\ref{area1_fit_result}.
From the figure, it can be just barely made out that $b_h$ increases with frequency.
It means that the structure of single pulses may evolve with frequency,
and this phenomenon is going to be analyzed in details in Section~\ref{sect:evolution}.

\begin{figure}[H]
  \centering
   \includegraphics[width=\linewidth]{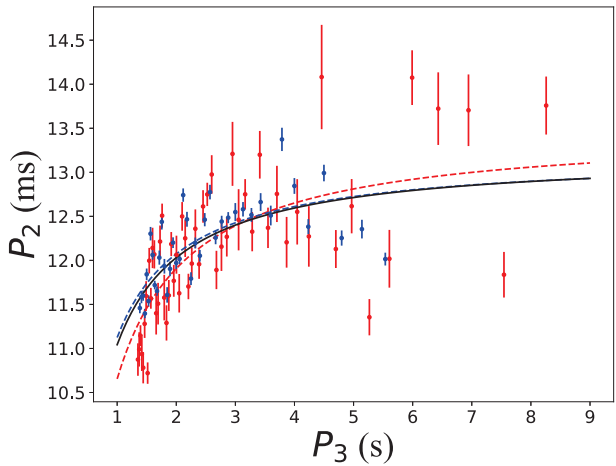}
   \caption{{\small The red points in Figure~\ref{drift_rate_f3}, blue points in
   Figure~\ref{harmonic_analysis} and corresponding fitting curves are plotted
   as the form of $P_2$ and $P_3$.
   The best-fitting result of all data points is shown as black curve.}}
   \label{p2_p3_relation}
\end{figure}

In fact, in the Eq.~\ref{simulatedf}, the intensity of frequency
$nf_0+f_2$ is proportional to $\exp\left(-\frac{2\pi^2n^2\sigma^2}{p_0^2}\right)$.
Obviously it reaches maximum while $n=0$, and in this case $nf_0+f_2=f_2$.
It means that, in Area 1, there is a good approximation $P_0\approx HP_2$.
With this approximation, Eq.~\ref{f2f3fit2} could be expressed as following,
\begin{equation}
P_2=\frac{P_3}{\gamma_h P_3+\beta_h},
\label{p2p3fit2}
\end{equation}
where $\gamma_h=75.8\pm0.8$ and $\beta_h=14\pm2$\,s.
Note that Eq.~\ref{p2p3fit2} has the same form as Eq.~\ref{p2p3fit1}.
Then we plot the $P_2$ and $P_3$ values obtained from Figure~\ref{drift_rate_f3}
and Figure~\ref{harmonic_analysis} as red and blue markers in Figure~\ref{p2_p3_relation},
and the fitting curves of Eq.~\ref{p2p3fit1} and Eq.~\ref{p2p3fit2}
are also plotted in the figure as red and blue dashed curves.
All the points in the figure are fitted with equation $P_2=\frac{P_3}{\gamma P_3+\beta}$.
The best-fitting result is shown as black curve with parameters
$\gamma=75.7\pm0.7$ and $\beta=15\pm1$\,s.

\begin{figure}[H]
  \centering
   \includegraphics[width=\linewidth]{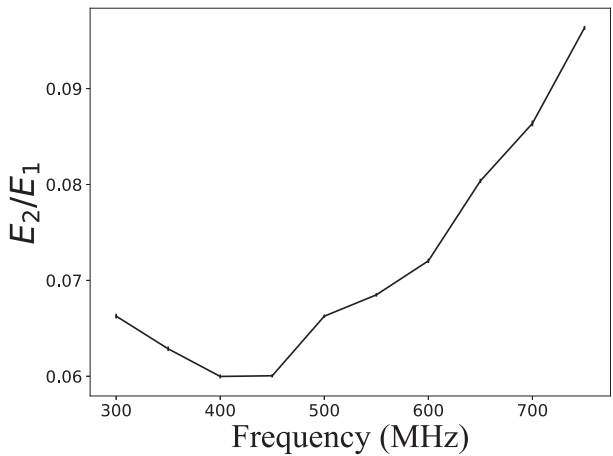}
   \caption{{\small The ratio of the intensity in Area 1 and Area 2 at different
   frequencies.}}
   \label{area2}
\end{figure}

In Figure~\ref{area1_fit_result}, the drift component in Area 2 is actually the ``slow''
drift mode component mentioned in Weltevrede, Edwards \& Stappers (2006) \cite{welt06} at 21\,cm,
and Weltevrede, Stappers \& Edwards (2007) \cite{welt07} pointed that this component vanished at $\sim$330\,MHz.
In Figure~\ref{area1_fit_result}, this component is much weaker than the fast drift
mode in Area 1.
We plot the ratio of the intensity in Area 1 and Area 2 in Figure~\ref{area2}.
From the figure, the ``slow'' drift component is stronger at high frequency than
at low frequency, and it is consistent with the result obtained in Weltevrede, Edwards \& Stappers (2006) \cite{welt06}.
It may also be the main reason why earlier observations have only seen this feature at
higher observing frequencies ($\sim$1.4\,GHz) and not at 300\,MHz~\cite{welt06,naid17,basu19}.

\section{Single Pulse: Correlation Scale Shrinking with Frequency}
\label{sect:evolution}

It is well known that the mean pulse profiles of most pulsars contract with frequency~\cite{thor91},
which is always explained by different radiation altitudes of radiation at different frequencies~\cite{kome70}.
However, the single pulse evolution with frequency is hard to study,
because of the weak flux and instable shape of individual pulse.
FAST can provide high signal-to-noise-ratio data of single pulse,
but the instable shape of single pulse is still a problem.
Additionally, the integrated pulse profile is the sum of single pulses, and it varies with frequency.
It means that we need to exclude the effects related to integrated pulse profile
while analyzing the single pulse evolution.
Here, a new method is practised for solving these problems.

\subsection{Single Pulse Structure Shrink with Frequency}
\label{sect:single_pulse_contract}

In a single pulse train composed of $n$ individual pulses at a certain frequency $\nu_1$,
the flux variation at a specified phase $\phi_1$ could be written as $I_{\nu_1,\,\phi_1}(n)$.
This radiation is generated in magnetosphere by the particles which originate from polar cap region.
Every bunch of particles will lead to a sub-pulse, which contains wideband radiation.
Naturally, the particle bunches which originate from the same region of polar cap will
generate emission with similar spectra, because of the same accelerating process.
Of course, the emission process in pulsars is indeed non-linear,
and the radiation spectrum would be different for particles with different energy.
But for analysis below, the spectrum at different pulse phases do not need to be exactly the same.
If only the radiation spectra of each bunch is monotonic along frequency,
the analysis could be taken.
As shown in Equation 4.3 of Melrose (1995) \cite{melr95}, this monotonicity
could be satisfied here.
Then, at another frequency $\nu_2$, there must be a phase $\phi_2$ at which the flux
variation series $I_{\nu_2,\,\phi_2}(n)$ is positive correlated to $I_{\nu_1,\,\phi_1}(n)$.
In this case, $\phi_2$ at frequency $\nu_2$ could be treated as the corresponding phase of
$\phi_1$ at frequency $\nu_1$.

\begin{figure}[H]
  \centering
   \includegraphics[width=\linewidth]{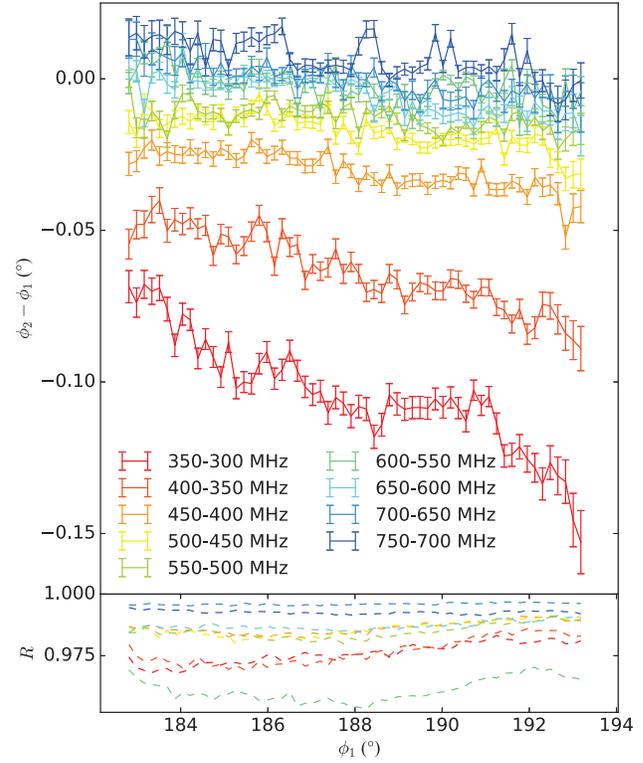}
   \caption{{\small The corresponding phase difference $\phi_2-\phi_1$ between adjacent narrow bands
   at different phase $\phi_1$ is shown in the top panel.
   The obviously down trending of each curve shows the shrinking of the single pulse structure.
   The maximum linearly dependent coefficient at different phase is shown in the bottom panel.}}
   \label{phase_contraction}
\end{figure}

In calculation, via maximizing the linearly dependent coefficient between $I_{\nu_2,\,\phi_2}(n)$ and $I_{\nu_1,\,\phi_1}(n)$, the phase $\phi_2$ could be determined.
However, in fact, for a specified phase $\phi_1$ at frequency $\nu_1$, the corresponding phase
$\phi_2$ at frequency $\nu_2$ may not exactly right at a phase data bin.
Then, the flux series between two adjacent phase bins at frequency $\nu_2$ used in analysis
is the linear interpolation values, and more exact $\phi_2$ could be obtained.
The relation between $\phi_1$ and $\phi_2$ is shown in Figure~\ref{phase_contraction}.
Considering the relation between $\phi_2$ and $\phi_1$ at the first order,
\begin{equation}
\phi_2=l_{\nu_2-\nu_1}\phi_1+d_{\nu_2-\nu_1},
\label{phirelation}
\end{equation}
where $l_{\nu_2-\nu_1}$ and $d_{\nu_2-\nu_1}$ are the linear slope and intercept to be fitted.
Obviously, $l_{\nu_2-\nu_1}$ is right the ratio of structure
width of single pulse at frequency $\nu_2$ and that at frequency $\nu_1$,
and is defined as correlation scale ratio between $\nu_2$ and $\nu_1$.
If single pulse structure doesn't evolution with frequency,
$l_{\nu_2-\nu_1}$ would be equal to 1 exactly.
If $l_{\nu_2-\nu_1}>1$, it means that the single pulse structure
(at the radiative interval, within pulse phase range $\sim180-195\degree$)
is wider at frequency $\nu_2$ than at frequency $\nu_1$.
In reverse, the single pulse structure contraction at frequency $\nu_2$ will lead to $l_{\nu_2-\nu_1}<1$.
From Figure~\ref{phase_contraction}, if $\nu_2>\nu_1$, $l_{\nu_2-\nu_1}$ is obviously less than 1 for PSR B2016$+$28.

For real data, the structure width of single pulses at different narrow bands with
respect to that at 300\,MHz is calculated.
In fact, to avoid the significant discrepancy of single pulse shape
due to large difference between frequencies, we just calculate the
structure width ratio between the single pulses at adjacent narrow bands.
Then the width ratio between two arbitrary narrow bands can be computed as the
cumulative product of the width ratio between adjacent narrow bands, e.g.,
$l_{500-350}=l_{500-450}\cdot l_{450-400}\cdot l_{400-350}$.
It should be noted that the absolute delay time between data at different frequencies
is not concerned in analysis above, and only the single pulse structure width
(i.e., the relative variation speed along pulse phase) is considered.
Thus, the time delay differences at different frequencies caused by
dispersion or other effects would not affect the result.
Also, because the pulse phase phi is just a relative value with respect
to the data at the same period, the flux variation effect at different periods, e.g. sub-pulse drifting, would not affect the analysis.
However, the flux variation within one single pulse,
e.g. quasi-periodic microstructure of PSR B2016+28 \cite{mitr15}, may affect.
In this case, the continuity of the phase differences versus frequency
and pulse phase need to be checked carefully.
Additionally, it should be noted that there is no factual value
for the single pulse width, and what could be obtained is the relative structure width (relative to that at a specific frequency) at different frequencies,
rather than the values of the single pulse width.

\begin{figure}[H]
  \centering
   \includegraphics[width=\linewidth]{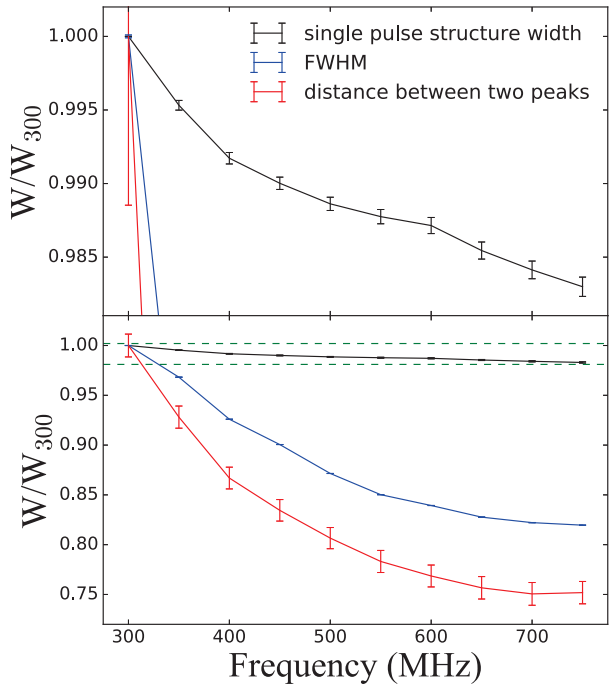}
   \caption{{\small The evolution of single pulse structure width (black markers) and mean pulse
   width (FWHM is denoted by blue markers and distance between components in two peaks fitting
   is denoted by red markers) with frequency is shown.
   Top panel is the magnified view of the part between two green dashed lines in bottom panel.}}
   \label{single_pulse_evolution}
\end{figure}

The evolution of single pulse structure width with frequency
is shown in Figure~\ref{single_pulse_evolution}.
To compare with width evolution of mean pulses, the full width at half maximum
(FWHM) of pulse profiles and the distance between two fitted square hyperbolic secant
components in the left panel of Figure~\ref{profile_fit} are also plotted.
In the bottom panel of Figure~\ref{single_pulse_evolution}, the part between two green dashed lines is also
zoomed in and shown in the top panel.
As shown in the figure, the mean pulse contracts by $\sim20\%$ from 300\,MHz to 750\,MHz,
while the single pulse structure only contracts by $\sim1.5\%$.
This contraction of single pulse structure consists with the slightly increasing $b_h$
shown in Figure~\ref{area1_fit_result}.
The contraction of single pulse structure implies that it is hard to align the pulse phase
at different frequencies.
Further more, the longitude resolved fluctuation spectrum may not be significant, especially at low frequency.

\subsection{Single Pulse Correlation Scale Shrinking in Dipole Field Model}

To explain the single pulse structure contraction,
the radiation geometry should be determined.

\begin{figure}[H]
  \centering
   \includegraphics[width=\linewidth]{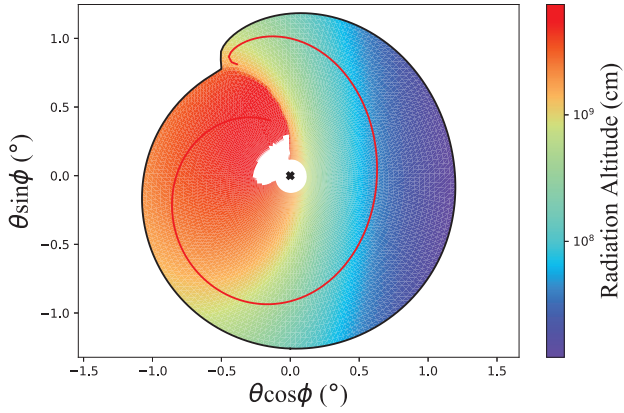}
   \caption{{\small The radiation altitude of every magnetic field line which
   passes polar cap region, and $\theta$ and $\phi$ are the zenith angle and
   position angle relative to magnetic axis.
   The black (red) curve denotes the intersection curve of pulsar surface and the
   last opening field lines (critical field lines).}}
   \label{radiation_altitude}
\end{figure}

Firstly, the magnetic field configuration should be obtained.
In this paper, the magnetic field of a rotating oblique dipole given in Cheng, Ruderman \& Zhang (2000) \cite{chen00} is adopted.
With the magnetic field, the magnetic field line could be calculated,
and the last opening field line (the field line which is tangent to the light cylinder)
and the critical field line (the field line which has a tangent line perpendicular to magnetic axis
at the intersection with light cylinder) could also be determined.
The region between the last opening field line and the pole is opening field line region (OFLR), and
is believed to be the emitting region~\cite{rude75}.
While the particles move along field line in OFLR, they will emit radiation.
The radiation direction would not be right along the magnetic field,
but deviates a little from the field because of the aberration effect~\cite{wang06}.
If the radiation direction is parallel to line of sight, the telescope could receive the radiation.
In fact, every field line in OFLR has no more than one point where the
radiation direction is parallel to line of sight before it intersects with the light cylinder.
Thus, for each field line in OFLR, the radiation position could be defined at this point,
and the corresponding radiation altitude could be calculated.
On the other hand, the field line in OFLR will intersect with pulsar surface at the polar cap region.
Then, each point in polar cap region corresponds to a radiation altitude,
and the result has been shown in Figure~\ref{radiation_altitude},
where $\theta$ and $\phi$ are the zenith angle and position angle relative to magnetic axis.
In the figure, the black (red) curve is the intersection curve of pulsar
surface and the last opening field lines (critical field lines).
From the figure, given the radiation at different frequencies originating from different altitudes,
the radiation positions must belongs to different magnetic field lines.

\begin{figure}[H]
  \centering
   \includegraphics[width=\linewidth]{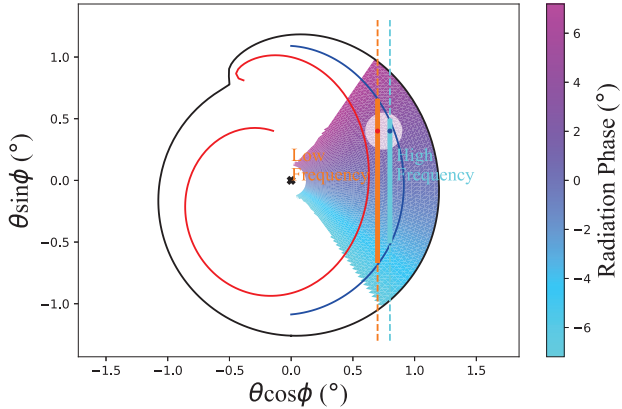}
   \caption{{\small As Figure~\ref{radiation_altitude}, but for radiation pulse phase.
   Assuming the radiation concentrates in the field lines which pass the blue curve,
   and the cyan (orange) line is the line group which radiate high (low) frequency emission.
   Thus, the solid cyan (orange) line segment denotes the width of mean pulse profile at high (low) frequency.
   Given a bunch of magnetic field lines at the white disc generate a sub-pulse,
   the cyan (orange) point would be the center of this sub-pulse at high (low) frequency.}}
   \label{radiation_phase}
\end{figure}

With the radiation altitude, the pulse phase corresponding to every point
in the polar cap region can be calculated.
In the calculation, the retardation effect~\cite{wang06}
has also been taken into account.
The results are shown in Figure~\ref{radiation_phase}.
In the figure, the pulse phase of point on the last opening field line at $\phi=0$
is set to be 0, and the region with pulse phase in range -7.2$\sim$7.2$\degree$ is plotted.

Then a phenomenological theory would be put up to explain the single pulse structure evolution.
For the case that the radiation is generated from the magnetosphere
between the last opening field line and the critical field line
(i.e., the annular region; the analysis for the core region between
the critical field line and the pole is similar), without loss of
generality, it could be assumed that the radiation field lines are
close to the field line group which pass the blue curve in Figure~\ref{radiation_phase}.
Besides, we also assume the field line groups which emit high and low frequency radiation,
and they intersect with the polar cap region at the cyan and orange lines, respectively
(for the actual situation, the intersection may be a complex curve).
Here the line corresponding to high frequency radiation is assumed further from magnetic
axis than that corresponding to low frequency radiation, for the pulse profile at high
frequency is narrower than at low frequency.
Accordingly, the line segments (marked as solid lines) between the intersection
points of these lines and blue curve could denote the pulse profile width.
In this scenario, if a bunch of magnetic field lines correspond to a sub-pulse
is at the region of white disc in Figure~\ref{radiation_phase}, the center of the sub-pulse at different
frequencies are marked as cyan and orange points.
Obviously, the pulse phase of this sub-pulse at different frequencies is different,
and the structure width of single pulse at low frequency would be a little wider.

In the phenomenological model above, the key assumption is the finite scale
of the region corresponding to a sub-pulse, and this scale is in a new dimension
perpendicular to the line of sight.
However, determining this scale needs accurate intersection line between
the radiative field lines and the polar cap (similar to the cyan and orange line in Figure~\ref{radiation_phase}).
To obtain this intersection, precisely polarization-calibrated data are necessary,
and it is expected to be achieved in the future.

\section{Discussions}
\label{sect:discussion}

\subsection{Mean Pulses}
\label{sect:discussion:mean}

The observational data used above are not calibrated,
and in some extent it would influence the analysis.
However, it could be estimated that the main results above are almost not affected.
The shape of the mean pulse profile would be influenced,
but the component properties would not change~\cite{lu16}.
Also, the uncalibrated data would impact on the shape of sub-pulse.
Whereas, this impact varies slow with time,
then the statistics of drifting phenomenon (e.g., the relation between $P_2$ and $P_3$) are not affected.
Besides, the rotation measure (RM) would also lead to flux fluctuation with frequency in the uncalibrated data.
The RM of PSR B2016$+$28 is -34.6\,rad\,m$^{-2}$~\cite{manc72},
which will result in a polarization position angle change of about 36\,rad from 275\,MHz to 775\,MHz.
It means that the RM would barely lead to monotonic changes with frequency
such as the mean pulse narrowing and correlation scale shrinking.

Lee et al. (2009) \cite{lee09} had also studied the profile evolution of PSR B2016$+$28,
and the result they obtained shows that the profile is wider at 1408\,MHz than that at 408\,MHz.
However, from Figure~\ref{single_pulse_evolution}, the profile contracts as frequency increasing between 275\,MHz and 775\,MHz.
This difference may be caused by three reasons:
(1) the trend of profile evolution may change between 775\,MHz and 1408\,MHz;
(2) the Gaussian function \cite{lee09} used to fit the profile may lead to different result;
(3) the equation \cite{lee09} used to calculate the profile width (Eq. 22) may result in a difference.

In our work, a conal beam model is used to fit the profiles,
and some geometric relations (e.g. Eq.~\ref{geometry} are adopted in the model.
However, these relations can only work for a perfect dipole field without considering
effects of abberation, retardation and magnetic field distortion.
In fact, the influence of these effects have been discussed in Lu et al. (2016) \cite{lu16},
and it has been shown that these effects could hardly make a difference for the profile fitting result.
On they other hand, we also assumed that the radiation cone has a shape of perfect round,
and this seems conflicting with Figure~\ref{radiation_phase}.
Actually, the conal beam model is used to describe the joint of two components,
and whether the cone shape is perfect round is unconsidered.
Besides, the radiation concentrating curve (blue curve in Figure~\ref{radiation_phase}) can also be treated as
circular arc, thus, to some extent, the round beam model is reasonable.

Additionally, in the paper, square hyperbolic function is not only used to fit the pulse profiles,
but also adopted for fitting the correlation result in Section~\ref{sect:corr_analysis}.
The square hyperbolic secant function is chosen because of its exponential wing.
Although square-hyperbolic-secant-like sub-pulse would not yield the same shaped
component of the cross-correlation result, the cross-correlation of two
square hyperbolic secant functions also have exponential wings.
In fact, cross-correlation of two $F_{\mathrm{sech~square}}(\phi,1,1,0)$ functions
is $2\left[2\phi\coth\left(\sqrt{\frac{2}{\pi}}\phi\right)-\sqrt{2\pi}\right]\csch^2
\left(\sqrt{\frac{2}{\pi}}\phi\right)$, whose curve is similar to that of
$F_{\mathrm{sech~square}}\left(\phi,\frac{2\sqrt{2\pi}}{3},\frac{3}{2},0\right)$.
This is the reason that the square hyperbolic secant function is still used to fit the
mean cross-correlation result.

\subsection{DM Correction}

The DM is modified in Section~\ref{sect:profile} by fitting the profile centric position.
However, DM of a pulsar is always changing, and it may even have a systemic changing rate $\dot{\mathrm{DM}}$.
Thus, the accurate of DM must be obtained via multi-observation analyzing.
On the other hand, it seems that there is still structured fitting residues in Figure~\ref{dm_fit}.
This may be caused by the effects of abberation, retardation and magnetic field distortion~\cite{wang06},
and radiation altitudes difference at different frequencies would also lead to this phenomenon.

Additionally, the roughly sinusoidal pattern sitting on top of the DM curve
is caused by the Faraday rotation effect.
As discussed in Section \ref{sect:discussion:mean}, the data are not calibrated.
Considering the Faraday rotation effect, at a certain pulse phase the polarization position angle (PPA) of the radiation will have a sinusoidal-like variation with frequency.
Noting the differential gain of the linear polarization feed,
the PPA curve versus pulse phase will lead to different total gain differences,
thus the observed intensity ratio of two peaks in the profile would deviate from the real value.
Then the fitted centric phase $\phi_0$ will be affected,
and the PPA evolution with frequency due to Faraday rotation effect will result in a
sinusoidal-like variation of $\phi_0$ with frequency.
With an RM of -34.6 rad$\cdot$m$^{-2}$, the variation of $\phi_0$ could just explain the roughly sinusoidal pattern.
It should also be noted that this variation is just caused by the differential gain, and the pulse does not intrinsically move along pulse phase.
Therefore, the single pulse analysis is not affected by this.

\subsection{Sub-pulse Drift Rate}

\begin{figure}[H]
  \centering
  \includegraphics[width=0.8\linewidth]{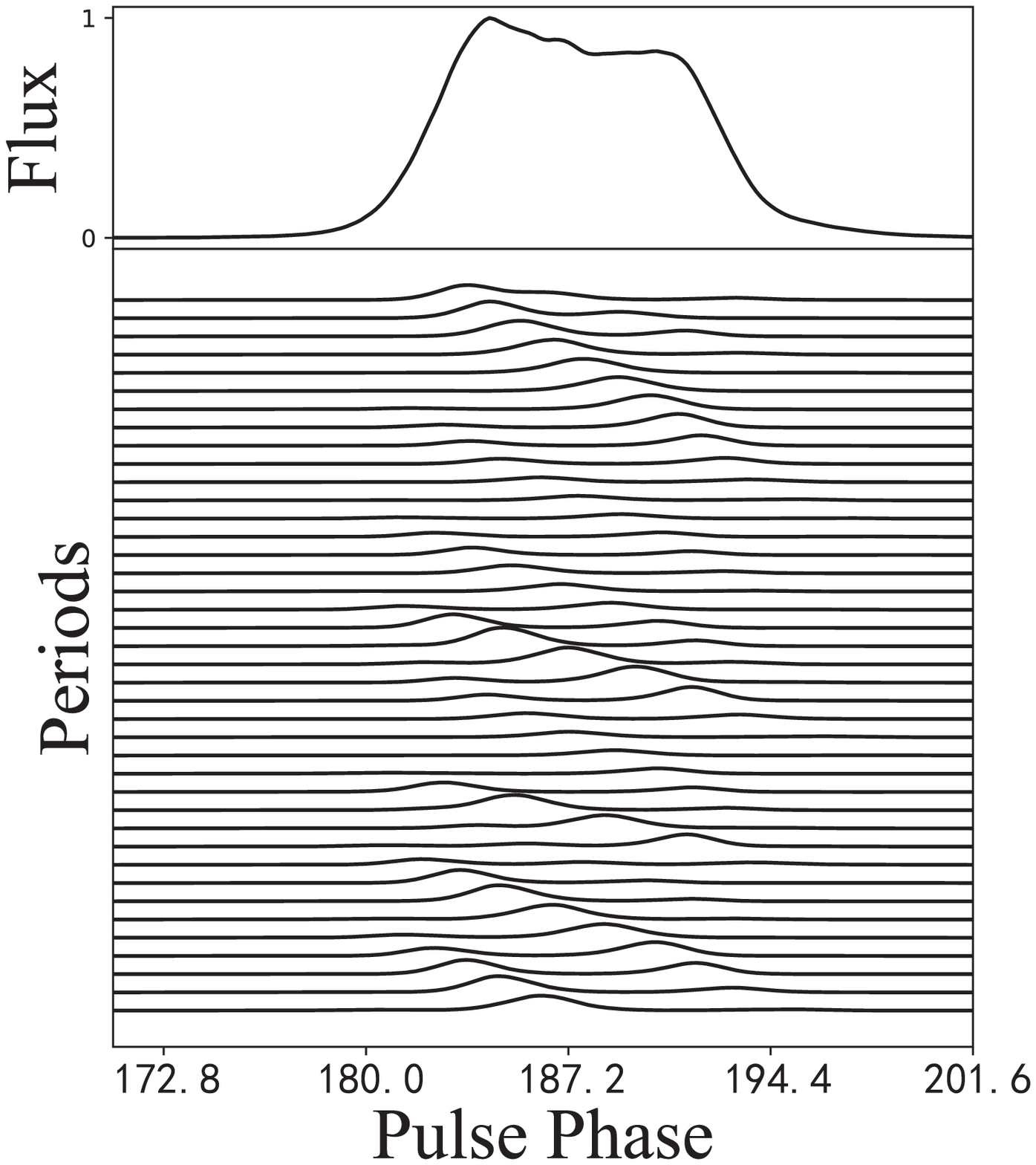}
  \includegraphics[width=1.0\linewidth]{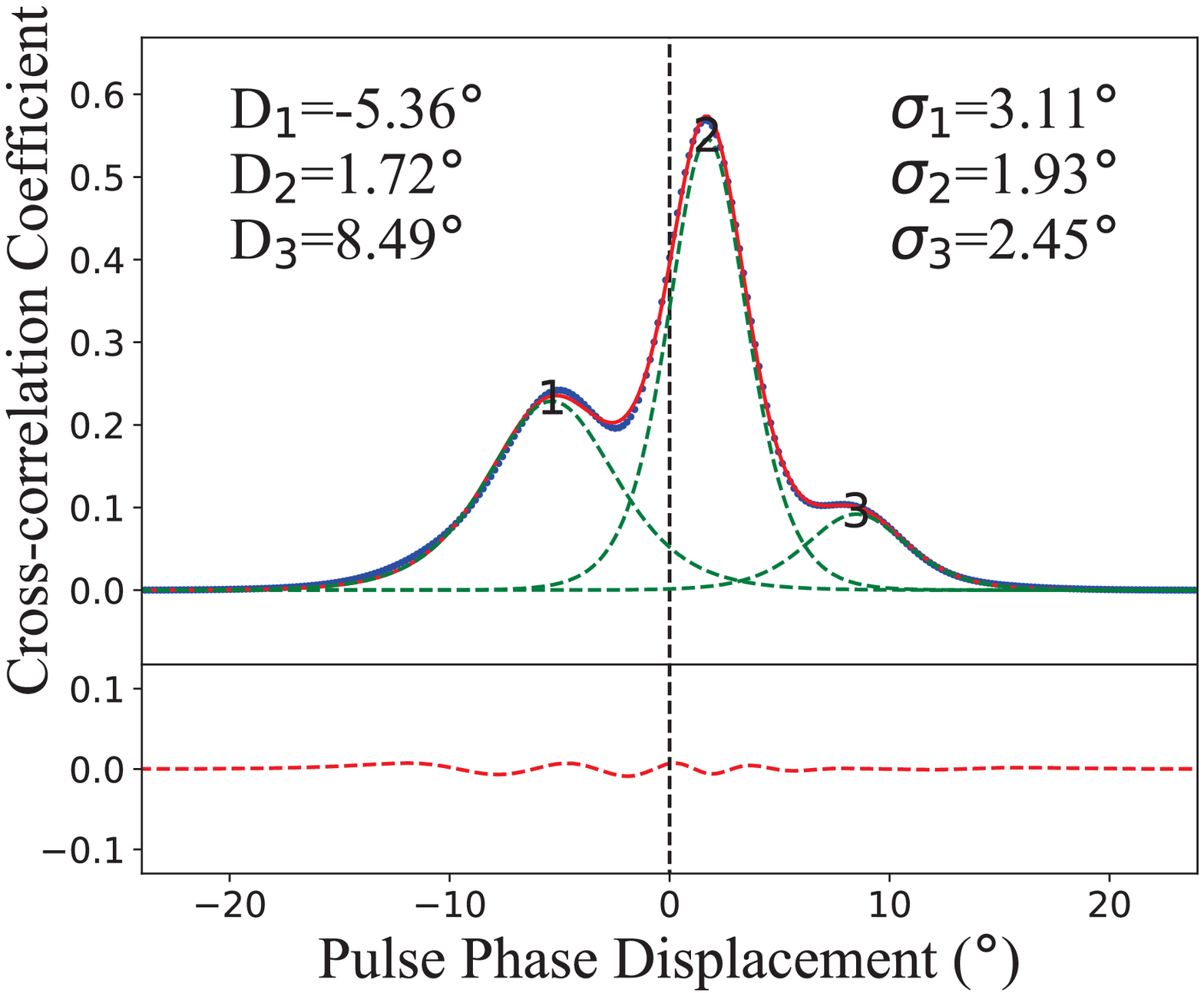}
   \caption{{\small The simulated single pulse train with a same profile of PSR B2016+28 at 300\,MHz are shown in the upper panel,
   and the cross-correlation results similar to Figure \ref{cross-correlation}.}}
   \label{simulation}
\end{figure}

The cross-correlation method was used in Section~\ref{sect:corr_analysis},
and the real drift band was then distinguished from the fake ones.
Although the theory had been put up in that section, its availability should be considered further.
To prove the validity of this method, a simulation was run.
In the upper panel of Figure~\ref{simulation}, the single pulse trains with drifting sub-pulses are simulated,
and the cross-correlation method was redone on them.
The correlation results are also shown in Figure~\ref{simulation}, which is similar to Figure~\ref{cross-correlation}.
The separation space between adjacent peaks is about $7.0\degree$
(which could be converted to $P_2\approx10.8$\,ms),
and it agrees with the input value 12\,ms in simulation.
Three peaks could be found in the figure, and the second peak is the narrowest with parameter $\sigma_2=1.93\degree$.
The position of this peak is $D_2=1.72\degree$, which means a drifting rate 1.72$\degree$/cycle.
With the input parameters $P_2\approx12$\,ms and $P_3\approx2$\,s,
the real drift rate should be $\sim2.16\degree$/cycle which is consistent with the fitting result.
Then the narrowest peak could be regarded as corresponding to the real drift band.
Additionally, parallel simulations were also run with different input parameters $P_2$ and $P_3$,
and similar results were obtained.
Therefore, the method used in Section~\ref{sect:corr_analysis} is proved to be feasible.

In Section~\ref{sect:drift_rate}, the drift rate of sub-pulse is measured to study the sub-pulse
instead of parameters $P_2$ and $P_3$.
This is because $P_2$ and $P_3$ could not be well-defined for a diffuse drifter,
but the drift rate is always stable for a specified sub-pulse.
In this case, the drift rate may be a more intrinsic quantity.
In fact, in the measurements process, empirical method is used to determine the drift
rate of each sub-pulse, the statistics may be inaccurate.

In Figure~\ref{harmonic_analysis}, the Area 2 is treated as the ``slow'' drift mode component mentioned in Weltevrede, Edwards \& Stappers (2006) \cite{welt06},
however, the drift rate of this component is not really slower than the component in Area 1.
Based on Figure~\ref{harmonic_analysis}, the drift rate of component in Area 2 could be estimated to be $\sim3\,\degree$/cycle.
Compared with Figure~\ref{drift_rate_f3}, this drift rate really could not be considered as ``slow''.

\subsection{Sub-pulse Shape}

In Section~\ref{sect:corr_analysis}, the distance between adjacent components in Figure~\ref{cross-correlation}
is referred to represent $P_2$.
Based on the value of distance $\sim6.1\,\degree$, $P_2$ could be estimated to be $\sim$9.5\,ms.
This value of $P_2$ is smaller than the results shown in Figure~\ref{p2_p3_relation}.
and this phenomenon may be caused by the variation of sub-pulse shape.
When we say sub-pulse drift, we mean that a sub-pulse moves to a new phase.
Whereas, the shape of a sub-pulse may also change.
Assume that the shape of sub-pulse is modulated by the mean pulse,
i.e., the received sub-pulse shape would be the product of a stable peak
at different pulse phase and the mean pulse profile.
Then the maximum point of a sub-pulse at the phase of profile wing would be
nearer to the profile center than the real radiation center.
In this case, the distance between adjacent components in correlation result
would give a smaller $P_2$ than the real one.
Besides, the radiation of sub-pulses may vary intrinsically, and this would
also affect the analysis of $P_2$
Also, the $P_2$ obtained by drift rate analysis and harmonic analysis would
be affected by this phenomenon.
But note that the drift rate is obtained via analyzing the data of multi-period (more than two),
and the harmonic analysis is even taken on the whole pulse trains.
Then the shape-variation phenomenon would affects little on them.

\subsection{Physics of Sub-pulse Drift}

In Ruderman \& Sutherland (1975) \cite{rude75}, the sub-pulse drift is ascribed to the drift of spark due to the
electric field difference $\Delta E$ between the real electric field component perpendicular to magnetic field and the rotation induced electric field.
Thus, the broad distribution of drift rate is caused by a diverse set of electric field difference.
Ruderman \& Sutherland (1975) \cite{rude75} (Eq. 30) showed that there is a positive relation between $\Delta E$
and the electric field component parallel to magnetic field $E_{\parallel}$.
Note that the sparks are caused by $E_{\parallel}$, therefore, it could be speculated
that the sparks would be sparser at lower $E_{\parallel}$ region, i.e., lower $\Delta E$ region.
Then the sparser the sparks are, the lower drift rate of sub-pulse should be.
On the other hand, $P_2$ is the distance between adjacent sub-pulses,
and could represent the angular distance between sparks.
Hence, the negative correlation between $P_2$ and drift rate (shown in Figure~\ref{drift_rate_f3}
and Figure~\ref{p2_p3_relation}) is consistent with the analysis above.
Additionally, if the sub-pulses with different drift rates correspond to different
region in polar gap, the pulse-profiles of them may be different.
However, no substantial change is observed as shown in Figure \ref{drift_rate_profiles}.

The diffuse drifting may be caused by the rough stellar surface.
The electric field around protuberances (i.e., small mountains) on stellar surface would be much higher than
the mean value, and could more likely lead to spark.
Then the spark could occur in irregular position of polar gap,
resulting in various drift rate.
The uneven surface could appear in the condensed surface of strangeon star~\cite{lu18}
or bare neutron star with high magnetic field~\cite{turo04}.
Also, diffuse drifting may be caused by the asymmetric shape of polar gap.
From Figure~\ref{radiation_altitude}, the intersection of last opening field lines and pulsar surface
is asymmetric, and the intersection of critical field lines and pulsar surface
is even not a closed curve.
This may result in an unclosed drift path of sparks, rather than the closed circuit
of PSR B0943$+$10~\cite{desh01}, and may also be the cause of diffuse drift.
More elaborate work on this subject is expected in the future.

\section{Summary}
\label{sect:summary}

To comprehend the electrodynamics of magnetosphere related to stellar surface properties,
a strong pulsar, B2016$+$28, is selected for FAST observation.
Both the mean and single pulses are extensively studied.

All the observational results suggest an RS-type inner vacuum gap on the polar cap of PSR B2016$+$28.
The mean pulse profiles may imply that the radiation is generated from an annular gap region,
in which sparks drift irregularly.
In this scenario, the pulse profile shrinking with frequency could be explained.
The correlation of drifting periods (positive relation between $P_2$ and $P_3$) could also
suggest a rough surface of pulsar, and the bonding energy of particles on stellar surface need
to be very high (e.g., in strangeon star or bare neutron star with extremely high magnetic field).
Single pulse structure width is also found to be shrinking with frequency, and it could be another
evidence of RS model.

FAST is the world¡¯s largest filled-aperture radio telescope,
and it is expected to show power in many fields~\cite{peng00a,peng00b}.
Now it is in the stage of commissioning and
early science and has limited capability to do sophisticated research.
Even so, the results above show the performance of
FAST to an extent, and it can be looked forward for FAST
to bring brilliant research output in the near future.

\Acknowledgements{
This work is supported by the National Key R\&D Program of China under grant number 2018YFA0404703 and 2017YFA0402602,
the National Natural Science Foundation of China (Grant No. 11673002 and 11225314),
the Open Project Program of the Key Laboratory of FAST, NAOC, Chinese Academy of Sciences,
and the project of Chinese Academy of Science (CAS) and the Max-Planck-Society (MPS) collaboration.
This work made use of the data from the FAST telescope (Five-hundred-meter Aperture
Spherical radio Telescope).
FAST is a Chinese national mega-science facility, built and operated by the
National Astronomical Observatories, Chinese Academy of Sciences.
The FAST FELLOWSHIP is supported by Special Funding for Advanced Users, budgeted and
administrated by Center for Astronomical Mega-Science, Chinese Academy of Sciences (CAMS).
YLY is supported by by National Key R\&D Program of China (2017YFA0402600)
and CAS ``Light of West China'' Program.
}

\InterestConflict{The authors declare that they have no conflict of interest.}




\section{}

\end{multicols}

\begin{thebibliography}{99}

\bibitem{lu18} J.~G. Lu, \& R. X. Xu, Quarks and Compact Stars 2017 (QCS2017), 011026 (2018)

\bibitem{rude75} M.~A. Ruderman \& P.~G. Sutherland, \apj, 196, 51 (1975)

\bibitem{gil03} J. Gil, G.~I. Melikidze \& U. Geppert, \aap, 407, 315 (2003)

\bibitem{xu99} R.~X. Xu, G.~J. Qiao \& B. Zhang, \apjl, 522, L109 (1999)

\bibitem{peng00a} B. Peng, R.~G. Strom, R. Nan, et al., Perspectives on Radio Astronomy: Science with Large Antenna Arrays, 
Proceedings of the Conference held at the Royal Netherlands Academy of Arts and Sciences in Amsterdam, 25 (2000)

\bibitem{peng00b} B. Peng, R. Nan \& Y. Su, Proceedings of SPIE - The International Society for Optical Engineering, 4015, 45 (2000)

\bibitem{lori95} D.~R. Lorimer, J.~A. Yates, A.~G. Lyne \& D.~M. Gould, \mnras, 273, 411 (1995)

\bibitem{hobb04} G. Hobbs, A.~G. Lyne, M. Kramer, C.~E. Martin, \& C. Jordan, \mnras, 353, 1311 (2004)

\bibitem{stov15} K. Stovall, P.~S. Ray, J. Blythe, et al., \apj, 808, 156 (2015)

\bibitem{oste77} L. Oster, D.~A. Hilton \& W. Sieber, \aap, 57, 1 (1977)

\bibitem{rank83} J.~M. Rankin, \apj, 274, 333 (1983)

\bibitem{welt06} P. Weltevrede, R.~T. Edwards \& B.~W. Stappers, \aap, 445, 243 (2006)

\bibitem{drak68} F.~D. Drake, \& H.~D. Craft, \nat, 220, 231 (1968)

\bibitem{tayl75} J.~H. Taylor, R.~N. Manchester \& G.~R. Huguenin, \apj, 195, 513 (1975)

\bibitem{rank86} J.~M. Rankin, \apj, 301, 901 (1986)

\bibitem{welt07} P. Weltevrede, B.~W. Stappers \& R.~T. Edwards, \aap, 469, 607 (2007)

\bibitem{naid17} A. Naidu, B.~C. Joshi, P.~K. Manoharan \& M.~A. KrishnaKumar, \aap, 604, A45 (2017)

\bibitem{jian19} P. Jiang, Y. L. Yue, H. Q. Gan, R. Yao, H. Li, G. F. Pan, J. H. Sun, D. J. Yu, H. F. Liu, N. Y. Tang, L. Qian, J. G. Lu, J. Yan, B. Peng, S. X. Zhang, Q. M. Wang, Q. Li, D. Li, and FAST Collaboration, Commissioning Progress of the FAST, Sci. China-Phys. Mech. Astron. 62, 959502 (2019).

\bibitem{hota04} A.~W. Hotan, W. van Straten \& R.~N. Manchester, \pasa, 21, 302 (2004)

\bibitem{edwa06} R.~T. Edwards, G.~B. Hobbs \& R.~N. Manchester, \mnras, 372, 1549 (2006)

\bibitem{hobb06} G.~B. Hobbs, R.~T. Edwards \& R.~N. Manchester, \mnras, 369, 655 (2006)

\bibitem{lu16} J.~G. Lu, Y.~J. Du, L.~F. Hao, et al., \apj, 816, 76 (2016)

\bibitem{lyne88} A.~G. Lyne \& R.~N. Manchester, \mnras, 234, 477 (1988)

\bibitem{kris83} S. Krishnamohan, \& G.~S. Downs, \apj, 265, 372 (1983)

\bibitem{mcki93} M.~M. McKinnon \& T.~H. Hankins, \aap, 269, 325 (1993)

\bibitem{shan12} R.~M. Shannon \& J.~M. Cordes, \apj, 761, 64 (2012)

\bibitem{ritc76} R.~T. Ritchings, \mnras, 176, 249 (1976)

\bibitem{sieb75} W. Sieber, \& L. Oster, \aap, 38, 325 (1975)

\bibitem{leeu02} A.~G.~J. van Leeuwen, M.~L.~A. Kouwenhoven, R. Ramachandran, J.~M. Rankin, \& B.~W. Stappers, \aap, 387, 169 (2002)

\bibitem{sery11} M. Serylak, B. Stappers \& P. Weltevrede, American Institute of Physics Conference Series, 1357, 131 (2011)

\bibitem{desh01} A.~A. Deshpande, \& J.~M. Rankin, \mnras, 322, 438 (2001)

\bibitem{basu19} R. Basu, D. Mitra, G.~I. Melikidze, \& A. Skrzypczak, \mnras, 482, 3757 (2019)

\bibitem{thor91} S.~E. Thorsett, \apj, 377, 263 (1991)

\bibitem{kome70} M.~M. Komesaroff, D. Morris \& D.~J. Cooke, \aplett, 5, 37 (1970)

\bibitem{melr95} D.~B. Melrose, Journal of Astrophysics and Astronomy, 16, 137 (1995)

\bibitem{mitr15} D. Mitra, M. Arjunwadkar, \& J. M. Rankin, \apj, 806, 236 (2015)

\bibitem{chen00} K.~S. Cheng, M. Ruderman, \& L. Zhang, \apj, 537, 964 (2000)

\bibitem{wang06} H.~G. Wang, G.~J. Qiao, R.~X. Xu \& Y. Liu, \mnras, 366, 945 (2006)

\bibitem{manc72} R.~N. Manchester, \apj, 172, 43 (1972)

\bibitem{lee09} K.~J. Lee, X.~H. Cui, H.~G. Wang, G.~J. Qiao \& R.~X. Xu, \apj, 703, 507 (2009)

\bibitem{xu06} R.~X. Xu, X.~H. Cui \& G.~J. Qiao, \cjaa, 6, 217 (2006)

\bibitem{turo04} R. Turolla, S. Zane, \& J.~J. Drake, \apj, 603, 265 (2004)

\end{thebibliography}
\end{document}